# The Structure of Space and Time, and the Indeterminacy of Classical Physics

Hanoch Ben-Yami[1]

ABSTRACT. I explain in what sense the structure of space and time is probably vague or *indefinite*, a notion I define. This leads to the mathematical representation of location in space and time by a vague interval. From this, a principle of complementary inaccuracy between spatial location and velocity is derived, and its relation to the Uncertainty Principle discussed. In addition, even if the laws of nature are deterministic, the behaviour of systems will be random to some degree. These and other considerations draw classical physics closer to Quantum Mechanics. An arrow of entropy is also derived, given an arrow of time. Lastly, chaos is given an additional, objective meaning.

> …it seems to me certain that we have to give up the notion of an absolute localization of the particles in a theoretical model. (Einstein 1934, 169)

## 1. INDEFINITE SEQUENCES

I start by describing a logical possibility, *an indefinite sequence*.

An *indefinite sequence* is a sequence of elements, each of which has a successor, but whose number of elements is bounded, smaller than some natural number.

> *Indefinite Sequence*: $\mathcal{S} = a_1, a_2, a_3 \ldots$; for each $a_i \in \mathcal{S}$, there is $a_{i+1} \in \mathcal{S}$; and for some $N$, for each $a_i \in \mathcal{S}$, $i < N$.

So, an indefinite sequence has no last element, but it is bounded.

Such a sequence is demonstrated by *sorites* sequences. If we start with a heap of sand and remove one grain after another, we end with grains which constitute no heap, so we have a bounded sequence of heaps. Yet the sequence of heaps has no last member: when a grain of sand is removed from a heap, the remaining collection is still a heap. The sequence of heaps is therefore an indefinite sequence. Similarly, if a baby grows older by a day, he is still a baby; but after a finite number of days, he is no longer a baby: for such a baby, the sequence 'baby *n*-days old' is therefore indefinite.

This idea might initially appear inconsistent, but any attempt I have come across to find an inconsistency fails. Usually, such attempts presuppose at some stage that indefinite sequences are impossible. And if sorites sequences are indeed indefinite sequences, then the idea is proved consistent by having instances.

[1] Department of Philosophy, Central European University, Vienna. benyamih@ceu.edu

– *But if each element has another one following it, why can't we add elements on and on until we have more than* $N$ *elements?* – This 'can' is an empirical one, not a logical one; and why an actual process should fail is not a question for logic. (In the case of sorites sequences it may fail because of the vague boundaries between a heap, say, and a collection which is not a heap.)

I have written more on this, and responded to some apparent difficulties, in (Ben-Yami 2010).

## 2. THE STRUCTURE OF SPACE AND TIME

We have no sufficient reason to hold that either space or time is *infinitely* divisible. This is not established by any empirical fact or conceptual considerations. If the only alternative to infinite divisibility were *finite* divisibility, then this would be conceptually problematic, if not contradictory, for familiar reasons. But once the possibility of *indefinite* divisibility is acknowledged, an alternative without these conceptual difficulties is available. Whether space or time are infinitely or indefinitely divisible should be decided by the consequences of each hypothesis and their agreement with observation.

What might it mean for the *distance* between two physical points, say, to be indefinitely divisible? Let us consider an elementary example (I have written on this in more detail, in Hebrew, in (Ben-Yami 2013, sec. 3.3)).

Suppose the length of a rigid rod is $L$. We can use it as a unit of measurement – our ruler – and measure the distance $D$ between points $A$ and $B$ as being, say, between $7L$ and $8L$. We next divide the rod into ten *roughly* equal units. In practice, we mark roughly equidistant notches on it. 'Roughly', since at this stage all we can say is that the new unit enters between the end of the rod more than 9 but less than 11 times – we have yet no more refined unit to improve on that determination. We again measure the distance between points $A$ and $B$ and get, for instance, the result that $7.4L < D < 7.5L$. Further divisions, by further notches or by other means, will help us improve the accuracy of the measurement. Perhaps we can determine that $7.472 < D < 7.481$ units. But the bounded accuracy in determining what counts as the end points of the rod and the edges of $A$ and $B$, as well as other factors, will make it impossible for us to proceed much beyond this accuracy.

Any non-finite sequence that gives the ratio $D/L$ with growing accuracy is indefinite. In this sense, the distance $D$ is *inaccurate*.

*These or related limitations will face us whatever we use as unit and whichever distances we measure*. Whatever is our standard $s$, and whichever is the distance $d$ we are measuring, the distance won't be determined much beyond some $\Delta$, such that $d = (x \pm \Delta)s$.

This $\Delta$ is *not* an expression of limitations of knowledge, but of bounded accuracy of distances in nature. It is not epistemic, but ontic.

$\Delta$ is also determined only to an indefinite degree of accuracy; that is why I used 'much beyond' in the formulation above.

I shall assume in this paper that the structure of space and of time is indefinite, and explore the implications of this assumption. I am interested primarily in the conceptual possibility, and I therefore won't try to add to my assumption specifications derived from contingent, factual physics laws. But I shall compare the conceptual results to contemporary physics, to show that the indefinite-structure assumption can explain a variety of physical phenomena. Some details of the specific application of the conceptual possibility to contemporary physics will depend on specifications coming from contemporary theories and should give more exact results, but this will not be attempted here. The success of these considerations in interpreting familiar physical phenomena shall support the applicability of this new approach to the structure of space and time.

## 3. THE REPRESENTATION OF LOCATION

I focus on location in space. However, what I write is also applicable with straightforward modifications to location in time, and below I shall consider time, or space and time together, where it is relevant.

In Newtonian physics and in Special Relativity, we represent the relative locations of bodies in space at a given moment by means of $\mathbf{R}^3$. In General Relativity the spatiotemporal manifold has additional complexities, which will not be addressed here, but the considerations below can be adapted to that more complex manifold. These considerations are applicable to configuration-space as well. Their applicability to phase space has some complications, which will become clear when we discuss complementary inaccuracy in Section 5.

If distances between bodies are indefinite, then the precision that $\mathbf{R}^3$ offers is beyond that found in nature. Given any unit of length, the distance between any two bodies relative to it is determined up to a vague, bounded degree of accuracy – our Δ above – while if we ascribe to each body, or to each body's centre of mass a point-location in $\mathbf{R}^3$, this distance is determined with infinite accuracy.

This does not mean that $\mathbf{R}^3$ cannot be used to represent physical space. It is not the only case in which this mathematical structure offers more structure than there exists in physical reality. Rotation, translation, and any linear transformation on $\mathbf{R}^3$ are also meaningless: if you add the same value to all representations of location in $\mathbf{R}^3$, say, you do not represent a different physical reality. Yet this redundancy in the system of representation does not block it from being usable in representing location in space, one only needs to secure that all facts about nature, the object of representation, will be invariant under translation in the representing structure.

Similarly, assuming the inaccuracy of distance, if we wish to represent the relative locations of bodies and the distances between them by means of $\mathbf{R}^3$, we should ascribe to each body a *vague region* of $\mathbf{R}^3$ as the representation of its location.

The relation between the vague region representing location and the indefinite structure of space and time can be explained as follows, with the help of Figure 1.

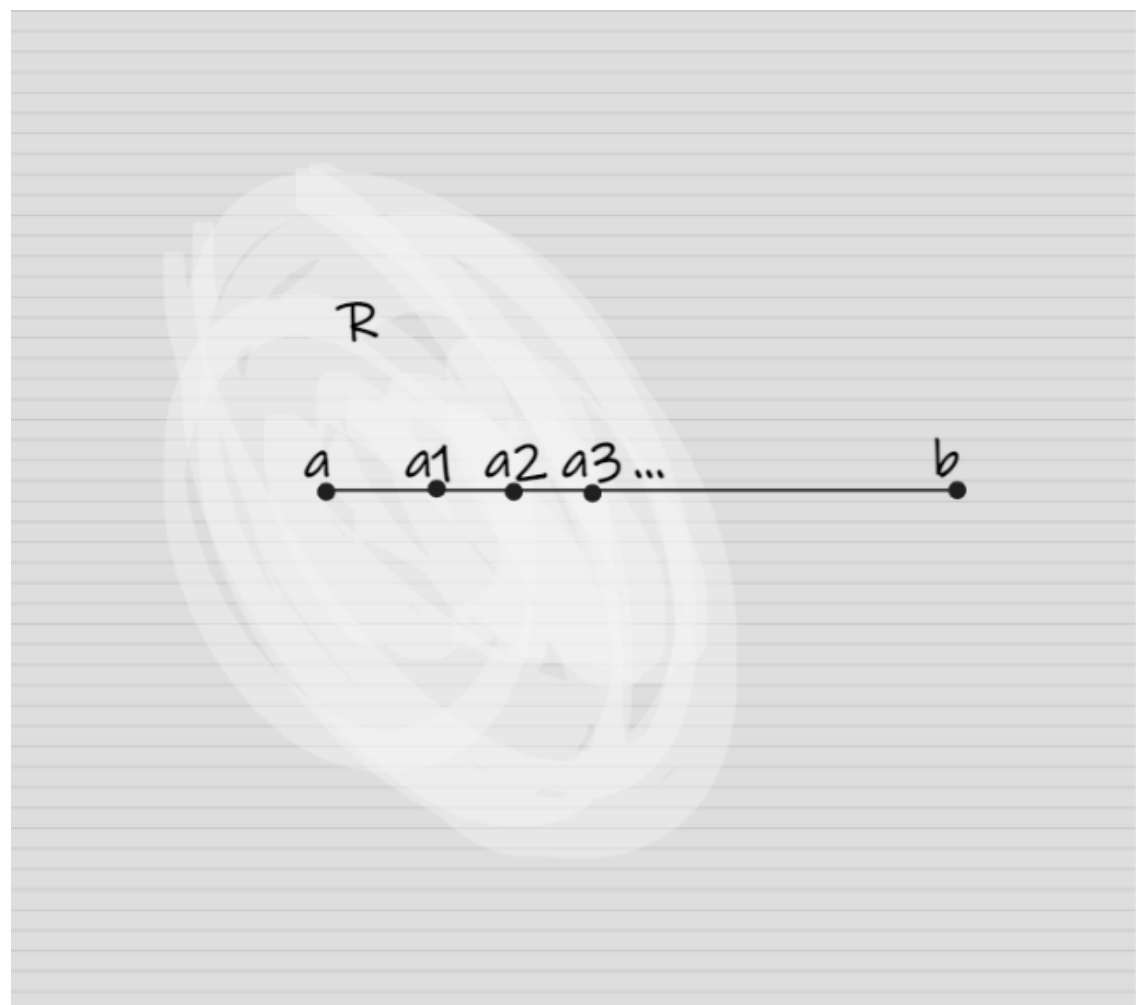

**Figure 1 Vague Region and indefiniteness**

Suppose $R$ is a vague region, and $ab$ a line with endpoints $a \in R$ and $b \notin R$. There's an indefinite sequence of points $a_1$, $a_2$, $a_3$ … such that $a_i \in ab$ and $a_i \in R$ for each $a_i$; for each $i$ and $j$, if $i < j$ then $aa_i < aa_j$; and if $c \in ab$ and $c \in R$ for some point $c$, then there is an $a_i$ such that $ac < aa_i$.

In practice, we can *approximate* such a representation by having a probability function, perhaps some Gaussian function, ascribe location to a body. The standard deviation or RMS of the Gaussian will approximate the degree of accuracy Δ in the determination of the body's location.

The fact that we can represent location only *approximately* in $\mathbf{R}^3$ is not a flaw in this representation, since location *is* determined only up to a bounded accuracy. But this does not mean that there are no better and worse approximations. Which method of mathematical representation will prove best is probably partly an empirical question.

The idea that the representation of location and of other physical quantities by real numbers is, in some sense, physically meaningless, is not new. It goes back at least to Max Born (Born 1955), who related it to a methodological principle dominant in modern physics and illustrated, for instance, by Einstein's Special and General Relativity:

> Statements like 'a quantity *x* has a completely definite value' (expressed by a real number and represented by a point in the mathematical continuum) seem to me to have no physical meaning. Modern physics has achieved its greatest successes by applying a principle of methodology, that concepts whose application requires distinctions that cannot in principle be observed, are meaningless and must be eliminated. (Born 1955, 81)

The question is, what should *replace* the representation of quantities by points in **R**, and specifically of location by points in $\mathbf{R}^3$? I don't think this has been addressed by Born.

An option suggested by Nicolas Gisin is to develop a new mathematics and a new concept of number – Gisin himself suggested replacing standard- by intuitionistic mathematics (Gisin 2021, n. 11; 2020). I find intuitionistic mathematics in the version Gisin adopts, which introduces temporal concepts into arithmetic, insufficiently clear. But more importantly, the current mathematical practice works sufficiently well, in the sense that it provides accurate predictions and does not omit any regularity of which we are aware in what it aims at representing. There is therefore no empirical need to modify it, and in fact these are good reasons for preserving it. What is needed is a different *interpretation* of the mathematical representation, and this is achieved by reading Δ, which we formerly took as a reflection of lack of knowledge alone, as a representation of the inaccuracy in location as well.

## 4. DETERMINATION OF THE INACCURACY

Although I have just argued for preserving the current mathematical representation of classical physics, this does not mean that there are no features of reality it does not represent, and that in this respect it cannot be complemented. (This need not require new mathematical apparatus; for instance, new conception of number.) If so, then classical physics can be considered *incomplete*. I shall now consider one such aspect.

We shouldn't think of the inaccuracy in location as provided by space and time independently of bodies' distribution in them. Space and time are not inert receptacles. Rather, the degree of inaccuracy is determined by distribution of bodies in space and their interactions. One possibility, for instance, is that the greater the number of bodies or their mass and the closer to each other they are, the lesser the inaccuracy in their vicinity. The more massive or energetic the bodies, the more fine-grained might space around them be. While if the bodies are remote and exert weak forces on each other – distant planets, for instance – then the distance between them is determined with lower accuracy. Moreover, as bodies interact and exchange energy, the kind of interaction – for instance how energetic it is – might also affect the degree of inaccuracy in their relative locations.

All these are abstract ideas about what might determine the inaccuracy in relative distances – abstract in the sense of not considering specific physical theories. The exact form of the dependence of inaccuracy on the interaction and distribution of bodies is to be determined not by such a priori reflections but by empirical research and by considerations on specific physical theories. (Theories of measurement might provide a clue here.) This will therefore not be attempted in this paper, which is mainly conceptual.

The inaccuracy in space-time location and the distributions of the bodies in it are to be determined together. This resembles the way it is in General Relativity, in which the curvature of space-time and the distribution of bodies in it are determined together.

## 5. A PRINCIPLE OF COMPLEMENTARY INACCURACY, AND "UNCERTAINTY"

Since distance and duration have a bounded degree of accuracy, so does velocity, as velocity is the ratio between the distance a body covered and the time it took to cover that distance.

Suppose a moving body's place is determined to a degree of accuracy of about $\Delta L$, which for simplicity we suppose does not change during the interval of its motion that we are considering. Accordingly, since both beginning and end locations are determined to a degree of accuracy of about $\Delta L$, the distance it travels is determined to a degree of accuracy of about $2\Delta L$. Let us designate it, $L \pm \Delta L$. Similarly, if the time a body is at a given location is determined to a degree of accuracy of about $\Delta T$, the time it takes to travel between beginning- and end locations is determined to a degree of accuracy of about $2\Delta T$. We designate it, $T \pm \Delta T$.

The velocity of the body between the two points is then,

$$v = (L \pm \Delta L) / (T \pm \Delta T)$$

Velocity is also determined to a bounded degree of accuracy. We shall say that velocity is *inaccurate*. This inaccuracy in velocity is *not* a matter of ignorance on our part, but it is an inaccuracy in nature.

We also see that *there is an inverse relation between the degree of accuracy of velocity and the interval relative to which that velocity is determined*.

Suppose a body travels with a roughly uniform velocity, in the sense that its average speed over large distances is constant. The smaller the interval $L \pm \Delta L$ we consider, the lesser the time T it takes the body to travel that distance, and the greater the relative inaccuracy 2ΔT/T in that time T. The inaccuracy in velocity, $\Delta v$, is therefore larger. And over very small distances L, in which ΔT approaches T, the inaccuracy approaches infinity. By contrast, for larger intervals L ± ΔL, T is larger and ΔT negligible relative to T, as is ΔL relative to L. The less the place of the body is determined, the more determinate is its average velocity over that interval. Location and velocity are complementary variables in their accuracy. (The complementary variables are not velocity and accuracy in location, but velocity and the *interval* over which this velocity is determined.)

I shall call this principle, *The Inaccuracy Principle*.

The Inaccuracy Principle is reminiscent of Quantum Mechanics' Uncertainty Principle (Heisenberg originally used *Ungenauigkeit* – *inaccuracy* or *imprecision* – and not *Unsicherheit* or *Ungewissheit*), but there are several issues confronting us if we wish to compare them. The way Heisenberg initially formulated the principle and argued for it (Heisenberg 1927) is not too clear and has been powerfully criticized, starting with Bohr and continuing to the present day; it might even simply be wrong (Ozawa 2003).[2] Later in the same year, Earle Hesse Kennard published a mathematical derivation of a 'preparation uncertainty principle', taking Heisenberg's 'measurement uncertainty principle' to be a special case of it (Kennard 1927). Heisenberg, in his 1930 lectures delivered at the University of Chicago, adopted Kennard's interpretation (Heisenberg 1930). But whether this relation between the principles indeed holds is debatable. However, while Heisenberg's principle, as argued for in (Heisenberg 1927), is controversial, for experimental reasons as well, Kennard's is a mathematical result of the physical theory and is unchallenged. In addition, there are several other formulations in the literature of related uncertainty principles, differing for instance in the way they define the uncertainty. Accordingly, when one talks about an Uncertainty Principle of Quantum Mechanics, which principle is meant should be clarified.

We shall consider the Uncertainty Principle (UCP) in its unproblematic and familiar Kennard's form. There are some significant differences between it and the Inaccuracy Principle. First, the corresponding complementary variables of the UCP are not *velocity* and place, but *momentum* and place. The UCP states, for instance, that $\Delta p_x \Delta x \geq h/4\pi$, while the Inaccuracy Principle conjugates $\Delta v_x$ and $\Delta x$. Secondly, the UCP specifies a quantity, $h/4\pi$, which serves as the minimum of the multiplication of the uncertainties or inaccuracies, while the Inaccuracy Principle is more qualitative in this respect, providing no quantitative minimum. Thirdly, the UCP does not limit the accuracy with which a specific variable can be measured, only the product of the two inaccuracies; $x$, for instance, can be measured according to it with arbitrarily high accuracy, $\Delta x$ approaching zero, accuracy which will be compensated by a growing uncertainty or inaccuracy in $p_x$. The *actual* inaccuracy in the measurement of $x$ will be due to the limitations of our instruments, according to it. The Inaccuracy Principle, by contrast, puts a lower limit on the accuracy of $x$, which is due to the indefinite structure of space-time; inaccuracy in measurements of $x$ is not due only to the limitations of our instruments. Lastly, the UCP is a result of an empirical theory, based on many non-obvious observations; it is contingent in nature. The Inaccuracy Principle, by contrast, is a result of very general reflections, mainly conceptual, on the nature of space and time.

Still, the Inaccuracy Principle, resulting from the indefinite structure of space-time, might be at the root of the Uncertainty Principle, although the latter adds some empirical, contingent determinations to the Inaccuracy Principle. The occurrence of momentum and not speed as the complementary variable of location in UCP might indicate that the distribution of the masses in space-time determines its indefinite structure, as was suggested in the previous section. An attempt to apply the ideas of the indefinite structure of space-time in Quantum Mechanics might also show how the inaccuracy in location and time are related to h/4π of the UCP.

[2] See (Hilgevoord and Uffink 2016; Atkinson and Peijnenburg 2022a; 2022b), to which I am indebted, for this and for other claims in this paragraph.

## 6. LAWS, INDETERMINACY AND DIRECTIONALITY

If the structure of space and of time is indefinite, then the location and velocity of a body at each moment – and a moment is not represented by a point on the line of real numbers but by a vague interval – these location and velocity are of bounded accuracy. Consequently, even if the laws of nature that describe the motion of bodies as a function of time are deterministic, there will be indeterminacy in the course of nature. (The laws can be deterministic in the sense that with the initial conditions of the system represented by *real* numbers they uniquely determine the future and past states of the system, again represented by real numbers, for any time.) This would make classical physics an *indeterministic* theory.

Suppose at a time $t_0 \pm \Delta t$ a body is in region $x_0 \pm \Delta x$ with a velocity $v_0 \pm \Delta v$. Then, given this inaccuracy in $x$ and $v$, the laws will determine a *range* of values for $x$ at any future time $t \pm \Delta t$. And as a rule, the larger is $t$, the greater the range which the laws will specify for $x$. However, the location of the body may continue to be inaccurate only up to the same $\Delta x$. Namely, there will be lower inaccuracy in place than the laws predict. This additional determination of location will be *random*, since it is not determined by any law.

We can illustrate it as follows. Suppose at $t_0 \pm \Delta t$ a body is at $x_0 \pm \Delta x$ with an average velocity $v \pm \Delta v$ (more on the velocity below). We assume the indeterminacies $\Delta t$ and $\Delta x$ do not change over time. The location in space-time of the body at any moment – a moment being a vague interval – will thus be represented by a spot of constant dimensions. Since the location is indefinite, the spot should have blurred boundaries, but this is not reproduced in our graph. We thus have the following representation of the possible change of place as a function of time:

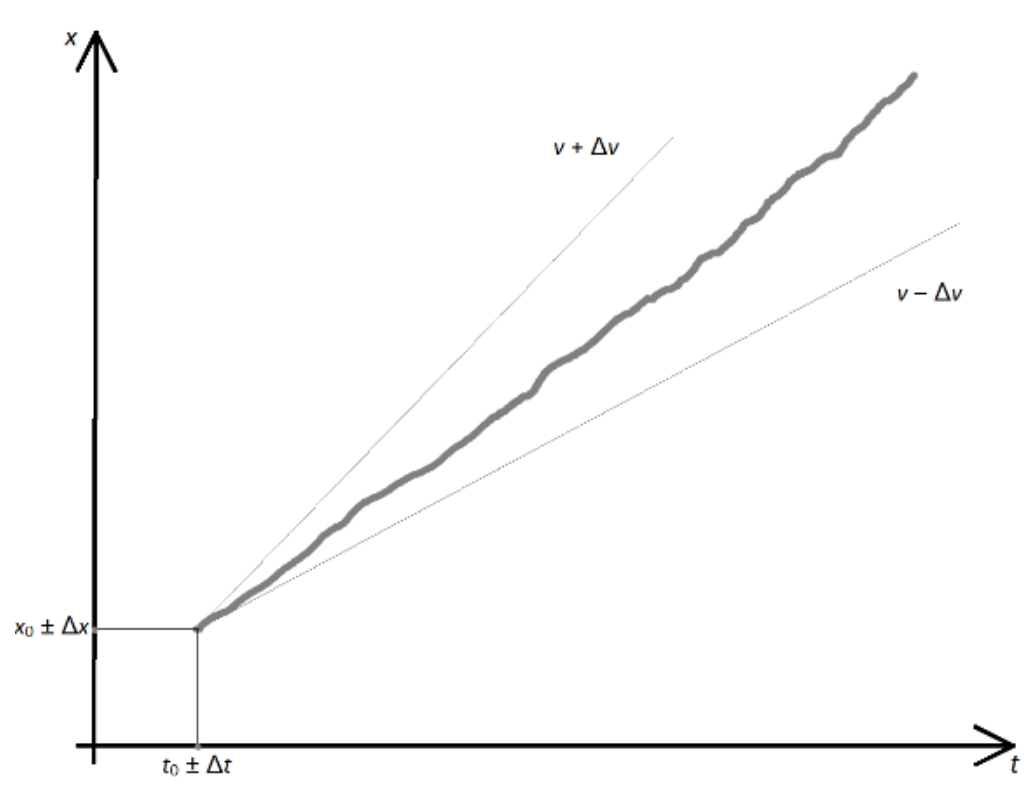

**Figure 2: Motion in Indefinite Space-time**

Given that the average velocity is $v \pm \Delta v$ (we are not here considering an accelerating body), the future locations as a function of time after $t_0 \pm \Delta t$ are within the area between $v + \Delta v$ and $v - \Delta v$, but the exact trajectory contains a random element. Moreover, I deliberately did not say that the *momentary* velocity is either $v$, $v \pm \Delta v$, or anything else, but that the *average* velocity is. The considerations in the previous section, on complementary inaccuracy, show the momentary velocity to be undefined. This can be illustrated if we zoom in on any segment of the graph above:

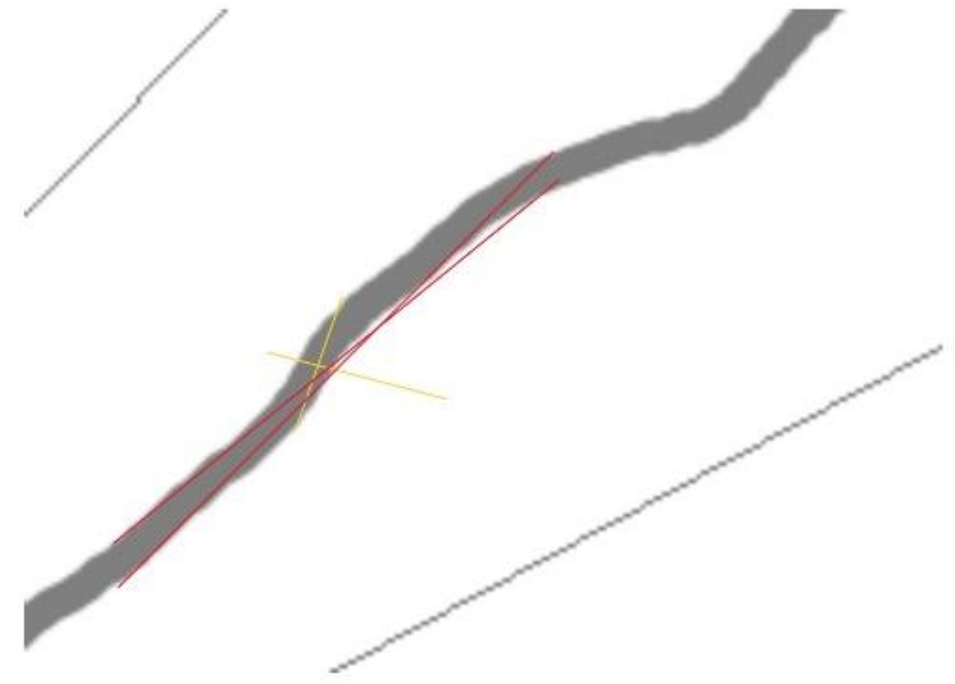

**Figure 3: Inaccuracy in Velocity**

The shorter the interval we examine, the greater the difference between maximal and minimal velocity, and the lesser the accuracy of the velocity the body has (yellow lines in Figure 3). Only over a large enough interval is the velocity defined with greater accuracy (red lines), converging asymptotically to a unique value as the interval grows.

If we represent the inaccurate location in space and time by a Gaussian function on $\mathbf{R}^3$, whose standard deviation approximates the degree of accuracy Δ in the determination of the body's location, and similarly for $v$, then the Gaussian assigns different probabilities to different locations and velocities, and this will show in the probabilities of future specific developments. The probability that if a body started at region *a* it will arrive after a time *t* at region *b* might be greater than that of its arriving at region *c*. Such a representation can give us a probabilistic distribution of the chances of the body being found at different regions. If such a representation can be added to the representation of location in space-time, then this will be another respect in which classical physics as currently conceived is incomplete.

These and earlier considerations draw classical physics closer to Quantum Mechanics.

Another consideration of this kind is the following. We assumed above, as is implicit in Figure 2, that given its initial location and velocity at time $t_0 \pm \Delta t$, namely $x_0 \pm \Delta x$ and $v \pm \Delta v$, the location of the body at *any* future time is specific up to degree of accuracy $\Delta x$, even if this location is only partly determined by the laws of nature given the initial conditions. However, if we allow for interaction – the exchange of energy with other bodies – to be intermittent, we can think of the trajectory of the body as follows. Given its initial conditions, the future location of the body becomes less and less determinate, according to the range and probabilities that the laws specify, until the body again interacts with other bodies at time $t_1$, and then its location relative to them is again determined up to degree of accuracy $\Delta x$. The probabilistic range that the laws determine for that time $t_1$ 'collapses' to the more accurate location, $x_1 \pm \Delta x$. We can graphically represent the process as follows:

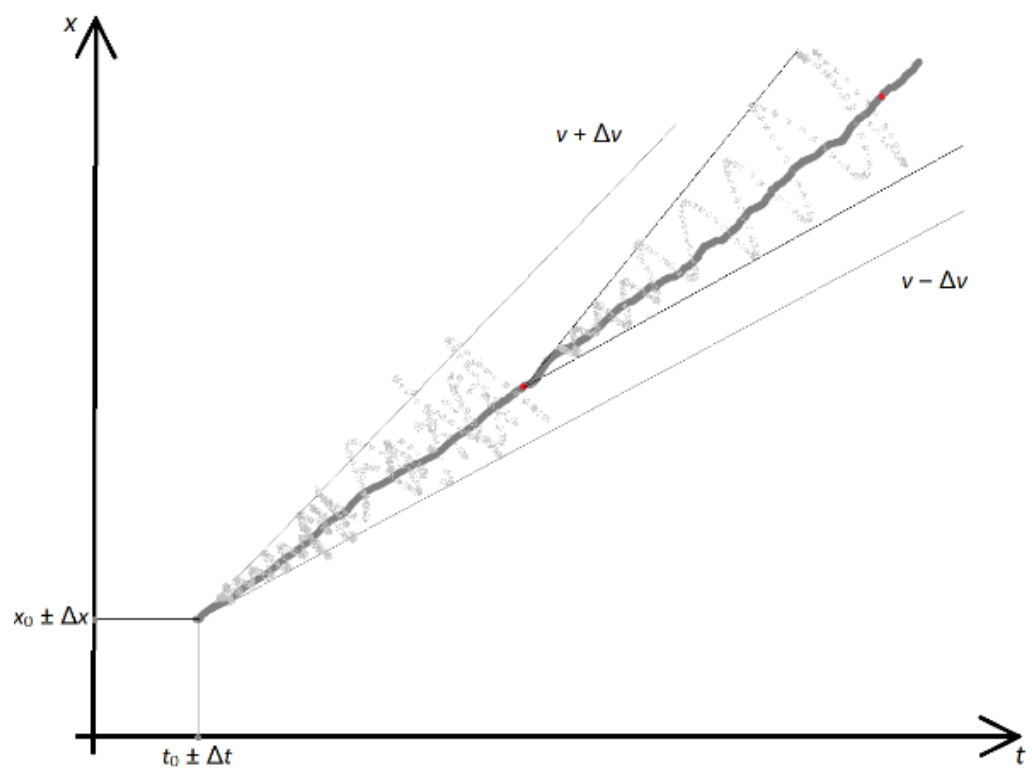

**Figure 4: Collapse in Classical Physics**

(Only the red spots represent actual location; the thicker dark grey line is left for comparison with Figure 2. The lighter-grey smudge represents the range of possible locations according to the laws between interactions or collapses.) Nothing determines to where exactly this collapse will lead, but given the probabilistic nature of the prediction of the laws, if many similar events occur, there will be a pattern in the collapse. This quantum-like behaviour is here derived with minimal assumptions on the actual laws of nature. The specific form that it will take is, however, to be determined according to the specific laws, a thing beyond what is attempted in this primarily conceptual paper.

## Entropy

If at a time $t_0$ bodies 1, … $n$ are in regions $x_i \pm \Delta x$ with velocities $v_i \pm \Delta v$, then the ranges of places and velocities that the laws determine at a later time $t$ include as a rule more possibilities with higher entropy than with lower one (this, I believe, is the typical situation with current laws of nature). Since the specific regions and velocities of these ranges that the bodies will have at $t$ are random (with the possibility of a probability distribution given by something like the option mentioned above), the entropy of the system will increase as a function of time. The arrow of time comes with an arrow of entropy.

These approach and result resemble those of coarse graining (e.g. (Robertson 2020)). However, the standard coarse graining approach does not see the coarse grain in location and velocity as a fact of nature. By contrast, the approach developed here, which assumes that the structure of space and time is indefinite, introduces *its* version of coarse grain as a means of the approximate representation in $\mathbf{R}^3$ or any other manifold of the indefiniteness of the quantities in nature.

Here and elsewhere, the indefinite structure of space and time turns what in other systems is lack of knowledge into lack of fact.

# 7. CHAOS

Chaos theory examines the behaviour over time of systems that obey deterministic laws. In a vast variety of systems, the behaviour of their elements displays sensitive dependence on initial conditions. Namely, in most cases, a difference in the initial conditions grows exponentially with time. If a body started its motion from a slightly different place or with a slightly different speed, then this slight difference would grow exponentially until very soon the body's place and speed would be far removed from what they actually were. And in case the range of magnitudes is bounded – for instance, if the body is confined to some volume, or its speed is bounded – then there will be no general functional dependence between the two results after some time. The place in which the body ends up after a time $t$ has nothing to do with the place it would have occupied after that time had it started its motion from a slightly different place, say. Such behaviour is called *chaotic*.

Our ability to predict the development over time of chaotic systems is limited. We always know the initial conditions of a system up to certain accuracy. If the system is chaotic, this accuracy will diminish exponentially with time. And if the range of the measured magnitudes is bounded, then after some time any possible result would be compatible with what we know of the initial conditions. Although the laws that describe the development of the system over time, as well as that development itself, are deterministic, we can in principle predict it with exponentially decreasing accuracy.

Chaos acquires an additional meaning if the structure of space and time is indefinite. Although the laws that describe a chaotic system's development over time are deterministic, the initial conditions of the system are fixed only up to a certain bounded accuracy. Since the place of a particle at a certain moment, say, is inaccurate, the particle's future behaviour can be any of the behaviours it would have had, had its place been accurate and compatible with its actual inaccurate place. This yields a range of possible behaviours, the difference between any two of which grows exponentially with time. Accordingly, the indefiniteness of space and time yields indeterminacy in the future state of the system, indeterminacy which grows exponentially with time. Moreover, in case the relevant magnitudes are bounded, after some time any possible state is compatible with any initial conditions.

We can therefore ascribe an additional meaning to what we understand by *chaos*. Suppose we are given a system with deterministic laws that is chaotic in the above defined sense. Suppose further that some of the magnitudes relative to which the system displays chaotic behaviour are inaccurate. Then the system is chaotic also in the sense that the indeterminacy in its future state, given specific initial conditions, grows exponentially with time. And if the relevant magnitudes are bounded, there is no general functional dependence between the state the system is in at a given time and the state it ends up in after some time. Chaotic indeterminacy is not only an essential feature of our knowledge, but is also a fact of nature.

## REFERENCES

Atkinson, David, and Jeanne Peijnenburg. 2022a. "Putting the Cart Before the Horse: Ernest Nagel and the Uncertainty Principle." In *Ernest Nagel: Philosophy of Science and the Fight for Clarity*, edited by Matthias Neuber and Adam Tamas Tuboly, 131–48. Logic, Epistemology, and the Unity of Science. Cham: Springer International Publishing. https://doi.org/10.1007/978-3-030-81010-8_7.

———. 2022b. "How Certain Is Heisenberg's Uncertainty Principle?" *HOPOS: The Journal of the International Society for the History of Philosophy of Science* 12 (1): 1–21. https://doi.org/10.1086/716930.

Ben-Yami, Hanoch. 2010. "A Wittgensteinian Solution to the Sorites." *Philosophical Investigations* 33 (3): 229–44. https://doi.org/10.1111/j.1467-9205.2010.01406.x.

———. 2013. *Aristotle's Hand: Five Philosophical Investigations*. Jerusalem: The Hebrew University Magnes Press.

Born, Max. 1955. "Is Classical Mechanics in Fact Deterministic?" In *Physics in My Generation*, 1969th ed., v:78–83. Heidelberg Science Library. Berlin, Heidelberg: Springer. https://link.springer.com/chapter/10.1007/978-3-662-25189-8_7.

Einstein, Albert. 1934. "On the Method of Theoretical Physics." *Philosophy of Science* 1 (2): 163–69. https://doi.org/10.1086/286316.

Gisin, Nicolas. 2020. "Mathematical Languages Shape Our Understanding of Time in Physics." *Nature Physics* 16 (2): 114–16. https://doi.org/10.1038/s41567-019-0748-5.

———. 2021. "Indeterminism in Physics, Classical Chaos and Bohmian Mechanics: Are Real Numbers Really Real?" *Erkenntnis* 86 (6): 1469–81. https://doi.org/10.1007/s10670-019-00165-8.

Heisenberg, Werner. 1927. "Über den anschaulichen Inhalt der quantentheoretischen Kinematik und Mechanik." *Zeitschrift für Physik* 43 (3): 172–98. https://doi.org/10.1007/BF01397280.

———. 1930. *The Physical Principles of the Quantum Theory*. Translated by Carl Eckart and Frank C. Hoyt. Chicago. University. Science Series. Chicago, Ill: The University of Chicago press.

Hilgevoord, Jan, and Jos Uffink. 2016. "The Uncertainty Principle." In *The Stanford Encyclopedia of Philosophy*, edited by Edward N. Zalta, Winter 2016 Edition. https://plato.stanford.edu/archives/win2016/entries/qt-uncertainty/.

Kennard, E. H. 1927. "Zur Quantenmechanik einfacher Bewegungstypen." *Zeitschrift für Physik* 44 (4): 326–52. https://doi.org/10.1007/BF01391200.

Ozawa, Masanao. 2003. "Universally Valid Reformulation of the Heisenberg Uncertainty Principle on Noise and Disturbance in Measurement." *Physical Review A* 67 (4): 042105. https://doi.org/10.1103/PhysRevA.67.042105.

Robertson, Katie. 2020. "Asymmetry, Abstraction, and Autonomy: Justifying Coarse-Graining in Statistical Mechanics." *The British Journal for the Philosophy of Science* 71 (2): 547–79. https://doi.org/10.1093/bjps/axy020.